# Anomalous pseudogap and superconducting state properties of heavily disordered $Y_{1-x}Ca_xBa_2(Cu_{1-y}Zn_y)_3O_{7-\delta}$

S.H. Naqib[1*], J.R. Cooper[1], R.S. Islam[1], and J.L. Tallon[2]

[1]IRC in Superconductivity, University in Cambridge, Madingley Road, Cambridge CB3 0HE, United Kingdom

[2]MacDiarmid Institute for Advanced Materials and Nanotechnology, Victoria University and Industrial Research Ltd, P.O. Box 31310, Lower Hutt, New Zealand

The role of substantial in-plane disorder (Zn) on the transport and AC susceptibility of $Y_{1-x}Ca_xBa_2(Cu_{1-y}Zn_y)_3O_{7-\delta}$ was investigated over a wide range of planar hole concentration, p. Resistivity, $\rho(T)$, for a number of overdoped to underdoped samples with $y \geq 0.055$ showed clear downturns at a characteristic temperature similar to that found at $T^*$ in Zn-free underdoped samples because of the presence of the pseudogap. Contrary to the widely observed behavior for underdoped cuprates at lower Zn contents (where the pseudogap energy increases almost linearly with decreasing p in the same way as for the Zn-free compounds), this apparent pseudogap temperature at high Zn content showed very little or no p-dependence. It also increases systematically with increasing Zn concentration in the $CuO_2$ planes. This anomalous behavior appears quite abruptly, e.g., samples with $y \leq 0.05$ exhibit the usual $T^*(p)$ behavior. AC susceptibility of these heavily disordered samples showed the superfluid density to be extremely low. Magneto-transport, $\rho(T,H)$, measurements are provisionally interpreted in terms of high-strength pinning centers for vortices in these samples. We also discuss various possible scenarios that might lead to a Zn induced pseudogap in the cuprates.

## INTRODUCTION

One of the most widely studied phenomena in the physics of high-temperature superconductors (HTS) is the "*pseudogap phase*"[1,2]. In the presence of a pseudogap (PG), various anomalies are observed in both normal and superconducting (SC) states, which can be interpreted in terms of a decrease in the quasiparticle density of states near the chemical potential[3]. Various theoretical models have been proposed to explain the origin of the PG[4-12]. But its nature remains an unresolved issue up to now.

In this paper we report systematic studies of the transport properties of $Y_{1-x}Ca_xBa_2(Cu_{1-y}Zn_y)_3O_{7-\delta}$. We have measured the DC resistivity, $\rho(T)$, room-temperature thermopower, S[290K], AC susceptibility (ACS), and magneto-resistivity, $\rho(T,H)$ of a series of sintered and c-axis oriented crystalline films of $Y_{1-x}Ca_xBa_2(Cu_{1-y}Zn_y)_3O_{7-\delta}$ compounds with different levels of Zn, Ca, and oxygen contents. The aim of present study was to examine the transport properties of $YBa_2Cu_3O_{7-\delta}$ (Y123) at high Zn concentration as a function of hole doping extending from highly overdoped (OD) to underdoped (UD) states. Pure Y123 with full oxygen loading ($\delta=0$), is only slightly OD, further overdoping is achieved by substituting $Y^{3+}$ by $Ca^{2+}$ [13,14]. The advantages of using Zn are: (i) It mainly substitutes the in-plane Cu(2) sites, thus the effects of planar impurity can be studied and (ii) the doping level remains nearly the same when Cu(2) is substituted by Zn enabling one to look at the effects of disorder at almost the same hole concentration[15,16]. Most of the studies on the effect of Zn on charge transport of Y123 are limited to the range of UD to optimum doping levels and at a moderate level of Zn substitution. Therefore, study of the system in the OD region with high levels of planar defects fills an important gap.

One interesting result of the present study is that, signs of the PG are seen clearly in overdoped $Y_{1-x}Ca_xBa_2(Cu_{1-y}Zn_y)_3O_{7-\delta}$ but only in a highly disordered ($y \geq 0.055$) state. This PG-like feature is almost independent of oxygen deficiency ($\delta$), indicating that it is induced by disorder and not by changing the hole concentration p.

## EXPERIMENTAL DETAILS AND RESULTS

Polycrystalline single-phase samples of $Y_{1-x}Ca_xBa_2(Cu_{1-y}Zn_y)_3O_{7-\delta}$ were synthesized by standard solid-state reaction methods using high-purity ( > 99.99%) powders. The details of sample preparation, characterization, and oxygen annealings can be found in refs. 13,14. High-quality c-axis oriented thin films of $Y_{1-x}Ca_xBa_2(Cu_{1-y}Zn_y)_3O_{7-\delta}$ were fabricated on $SrTiO_3$ substrates using pulsed laser deposition (PLD). Details of PLD, characterization, and oxygenation of the films can be found in refs.14,17.

Various normal and SC state properties including $T^*$ (the PG temperature) are highly sensitive to p and, therefore, it is important to determine p as accurately as possible. We have used the room-temperature thermopower, S[290K], which varies systematically with p for various HTS over the entire doping range extending from very underdoped to heavily overdoped regimes[18]. Previous studies, following the findings of Obertelli *et al.* [18], showed that S[290K] does not vary significantly with Zn in Y123 for $\delta < 0.5$ [16,19]. Accordingly, in this doping range S[290K] is still a good measure of p even in the presence of strong in-plane potential scattering by $Zn^{2+}$ ions. We have also calculated p for all the samples using a generalized $T_c(p)$ relation given by[14,19,20]:

$$\frac{T_c(p)}{T_c(p_{opt})} = 1 - Z(p - p_{opt})^2 \qquad (1)$$

For the Zn-free samples, Z and $p_{opt}$ take the universal values of 82.6 and 0.16 respectively[20], but these parameters increase systematically with increasing in-plane disorder[14,19]. Using p-values from S[290K] and the experimental values of $T_c$, a very good fit of $T_c(p)$ of the form of eq.(1) was obtained for all the samples. Fig.1 shows the case for $Y_{0.80}Ca_{0.20}Ba_2(Cu_{1-y}Zn_y)_3O_{7-\delta}$ sintered compounds. There is a systematic shift of $p_{opt}$ towards higher values (from $p_{opt} = 0.16$ for the impurity-free case) and an increase in Z with increasing Zn content. These findings have important consequences. Increasing Z(y) implies the shrinkage of the p-range over which the compound is superconducting upon Zn substitution, and the shift of $p_{opt}(y)$ towards higher values implies that the superconducting dome is displaced asymmetrically towards the overdoped side. This latter feature, we believe, is indicative of a fundamental difference between the underdoped and the overdoped regions, namely the existence of the pseudogap in the underdoped region and its absence in the overdoped region[21]. From Fig.1 it is also seen that SC is at its strongest at p ~ 0.186, as this remains the last point of SC at a critical Zn concentration (defined as the highest possible Zn concentration for which SC just survives considering all the possible doping states). This has been reported earlier in other studies and the value p ~

0.19 is indeed a special value at which the PG vanishes quite abruptly [4,14,19,21-24].

$T_c$ was obtained from both resistivity and low field ($H_{rms} = 0.1$ Oe; f = 333.3 Hz) ACS data. $T_c$ was identified at zero resistivity (within the noise level of $\pm 10^{-6}$ $\Omega$) and at the point where the line drawn on the steepest part of the diamagnetic ACS curve meets the T-independent base line associated with the negligibly small normal state (NS) signal. $T_c$-values obtained from these methods agree within 1K for most of the samples[19].

Patterned thin films with evaporated gold contact pads and high density (85% to 95% of the theoretical density) sintered bars were used for resistivity measurements. Resistivity was measured using the four-terminal configuration. In previous studies[4,17,19,24] we have analysed $\rho(T,p)$ of a large number of $Y_{1-x}Ca_xBa_2(Cu_{1-y}Zn_y)_3O_{7-\delta}$ sintered and c-axis oriented thin films with $y \leq 0.05$ and found the pseudogap temperature, $T^*(p)$, to be independent of Zn content and independent of the crystalline state of the compound (which also implies that $T^*(p)$ does not depend on the properties of the grain boundaries). We have also found clear indications that $T^*(p)$ vanishes at p = 0.19 $\pm$ 0.01 in these samples from our magnetic and transport measurements [14,19,24], consistent with the findings from the specific heat measurements by Loram et al.[3,23].

$\rho(T,p)$ of two heavily disordered $Y_{0.80}Ca_{0.20}Ba_2(Cu_{1-y}Zn_y)_3O_{7-\delta}$ sintered samples is shown in Fig.2. As is standard practice[2,25], we have taken $T^*$ as the temperature at which resistivity starts to fall at a faster rate with temperature than its high-temperature T-linear behavior[25]. It is clearly seen from Fig.2a that the evolution of $\rho(T,p)$ for the first y = 0.06 sample (S1) is different compared with other $Y_{1-x}Ca_xBa_2(Cu_{1-y}Zn_y)_3O_{7-\delta}$ compounds with y < 0.05 (shown in Figs.3). In this (S1) sample, there is a very large residual resistivity, also the maximum $T_c$ has been suppressed by ~ 57.1 K from its Zn-free maximum value (~ 85K), indicating that a large amount of Zn resides in the $CuO_2$ planes. Even at fairly high hole concentration, (p ~ 0.180) $\rho(T)$ is non-metallic over the whole experimental temperature range from 300K, indicating significant carrier localization induced by Zn in the $CuO_2$ plane and also due to some extrinsic effects like the degree of porosity of the sintered sample. The most striking feature of the $\rho(T,p)$ data is the downturn in resistivity at around 230 ± 5K, almost independent of planar hole concentration p. This downturn is somewhat masked for samples with higher oxygen deficiency (smaller p) due to the stronger non-metallic behavior over the whole temperature range. These downturns in the $\rho(T)$ data are similar to those observed in samples with a PG (see Figs.3). But in this sample it is unusual because (i) the PG-like feature is observed in resistivity for a heavily overdoped sample ($T_c$ = 12K, whereas $T_{cmax}$ ~ 27.9K for y = 0.06) and (ii) $T^*$ does not seem to change with p. The former indicates the possibility of disorder giving rise to this PG-like phenomenon, whereas the latter shows that in such a case, this feature is almost independent of planar hole content, p. In Fig.2b we have shown the p-independent $T^*$ for another heavily disordered compound (S2) from the $\rho(T)$ measurements. Although the nominal composition for S2 from a different batch was also y = 0.06 and x = 0.20, from the value of $T_{cmax}$ (= 32K) we think the actual Zn concentration in the CuO$_2$ plane is a little lower, probably ~ 5.5%.

$T^*$ for this sample is ~ 205 ± 5K. Several $\rho(T)$ measurements were done for these samples after each oxygen annealing and measurements were also repeated after time intervals of several months. Each time we have observed the characteristic downturn in $\rho(T)$ at the same temperature. It is worth mentioning that for a given value of p, $\rho(T,p)$ of S1 appears to be more non-metallic than S2. We believe this is not intrinsic, but that it is related to the different level of porosity of the samples, and likely to be governed by the difference in their densities[26,27]. The density of S2 (5.93 gm/cc) was larger than that of S1 (5.61 gm/cc). This is also supported from the values of $T_{cmax}$ of the samples that are well explained by the small difference in the actual levels of Zn in the CuO$_2$ planes. Extrinsic properties, like the quality of the conductivity at grain boundaries, can have a large effect on $\rho(T)$ without having a noticeable effect on $T_c$[17,27].

To explore the matter further, we have searched the existing literature extensively to find published transport data on heavily Zn-substituted samples as a function of carrier concentration. Data with more than 5%Zn is rare, and when available, $\rho(T)$ measurements were often not done over a wide enough range of hole concentrations. We have found one source of published data[28] where $\rho(T)$ was measured over a wide range of p-values for samples with y = 0.07. In that paper measurements were performed on high-quality c-axis oriented thin films of $YBa_2(Cu_{1-y}Zn_y)_3O_{7-\delta}$. We show the in-plane resistivity, $\rho_{ab}(T)$[28], for $YBa_2(Cu_{0.93}Zn_{0.07})_3O_{7-\delta}$ in Fig.2c. A p-independent $T^*$ is seen again at 250 ± 5K. From the reported[28] values of oxygen deficiencies for this film, we estimate the hole concentrations to be in the range p = 0.17 to 0.11 (within ± 0.01). At this point it is important to notice that resistivity of this crystalline thin film is not affected by the presence of significant grain boundary resistance. This illustrates the role played by Zn as the principal source of disorder giving rise to the appearance of this PG-like feature in these samples. It also appears that $T^*$ in these heavily disordered samples is determined by the amount of Zn in the $CuO_2$ plane. For example, we have $T^* = 205 ± 5K$, 230 ± 5K, and 250 ± 5K for samples with ~ 5.5%Zn, 6%Zn, and 7%Zn respectively. However, $\rho(T)$ of 4%Zn substituted samples shows the usual $T^*(p)$ behavior as was found for the Zn-free samples (see Fig.3b). This indicates that the presence of a disorder induced PG temperature is a non-linear function of Zn concentration and, at least in the $\rho(T)$ measurements, it reveals itself only in the highly disordered state. We discuss the possible implications and origins of this anomalous PG in the next section.

The AC susceptibility (ACS) (expressed in μV/gm) for both S1 and S2 is shown as a function of p and magnetic field in Fig.4. Once again quite different features are seen for these two samples compared with other sintered $Y_{1-x}Ca_xBa_2(Cu_{1-y}Zn_y)_3O_{7-\delta}$ samples with y ≤ 0.05 (see Fig.5). First, the magnitude of the low-field ACS signal at low-T is largely reduced for the y > 0.05 samples and shows a strong p-dependence, namely the ACS signal reduces rapidly with the decrease of hole content in the UD side. Both these features are absent for other samples for up to 5%Zn concentration (see Figs.5a and 5b). Second, the field dependent ACS signals as a function of temperature show the same qualitative features for both the sintered pellet and powder for S1 (see

Figs.4c and 4d), showing the absence of any coupling $T_c$ and also the absence of any intergrain contribution to the shielding current. The reduction in the normalized ACS signal in powdered sample simply implies that grain size is further reduced due to grinding. The rapid reduction of the low-field ACS signal with increasing underdoping in S1 and S2 suggests that the superconducting volume fraction decreases rapidly with decreasing hole concentration. Whereas for a less disordered sample (Fig.5d), the field dependent ACS clearly shows the intergrain shielding component and the grain-coupling temperature ($T_{coupling}$)[29].

Zn is believed to destroy superconductivity and superfluid density on a local scale (of the order of in-plane superconducting coherence length, $\xi_{ab}$)[30]. This seems to agree with ACS data for up to 5%Zn substitution where the low-field bulk ACS signal showed little variation with Zn and the entire volume of the sample contributes to the ACS signal at low-temperature. It is possible that for heavily Zn-substituted samples the effect is not local any more and some kind of phase segregation (between superconducting and non-superconducting regions) over a

*macroscopic* length scale takes place. Alternatively, the superfluid density in these highly disordered samples can be extremely low[31]. At temperatures near $T_c$, the temperature derivative of the ACS signal for the powdered samples can be used to estimate the zero-temperature superfluid density, $n_{s0}$. The temperature derivative of the ACS signal can be used to represent the quantity $N[a^2 n_{s0}]/[15T_c]$[32], where $N$ is a constant depending on the shape of the grains, and $a$ denotes the average grain size. In our calculations we have assumed $N$ and $a$ to be the same for all samples under consideration. This is a reasonable approximation because these compounds had nearly identical density and grinding of the pellets was done in exactly the same manner. We have calculated the fractional suppression in $n_{s0}$ as a function of Zn content for 3%Zn and 6%Zn substituted samples with p = 0.20 ± 0.005 (therefore, similarly heat treated). The results are as follows, $[n_{s0}/T_c]_{y=0.03}/[n_{s0}/T_c]_{y=0.0} = 0.48$ and $[n_{s0}/T_c]_{y=0.06}/[n_{s0}/T_c]_{y=0.0} = 0.07$. Thus, $n_{s0}/T_c$ for the 6%Zn sample is reduced to just 7% of its value for the Zn-free case at the same hole content. Whereas, $n_{s0}/T_c$ for the 3%Zn sample is reduced to 48% of its value for the Zn-free case. The above findings agree quite well with those obtained by Panagopoulos *et al.*[31] from their London penetration depth measurements on magnetically aligned powders (grains).

Next, we present $\rho(T,H)$ results in Fig.6 for some sintered and thin film $Y_{1-x}Ca_xBa_2(Cu_{1-y}Zn_y)_3O_{7-\delta}$ samples. Details of magneto-transport measurements can be found elsewhere[14,24]. The field broadening of the resistive transition is shown in Figs.6a and 6b for the 3%Zn substituted sintered and 2%Zn substituted thin film samples respectively. The difference between the values of $T_{c-onset}$ at 0 Tesla and 12 Tesla is ~ $(4.5 \pm 1)$K (see Fig.6a) but the difference in $T_c$ defined at the zero resistivity under these two applied fields is ~ $(18 \pm 1)$K for the sintered sample. For the thin film with magnetic field parallel to the c-axis, these values are ~ $(6 \pm 1)$K and ~ $(20 \pm 1)$K respectively. Such field broadening of the resistive transition is a general property for cuprates and one possible explanation is in terms of the unconventional vortex dynamics for these compounds[33]. Interestingly, the field broadening for the y = 0.06 sintered sample (S1) is insignificant even though the resistive transition itself is quite broad. For example, the difference between the values of $T_{c-onset}$ at 0 Tesla and 12 Tesla is $(9.5 \pm 1)$K and the difference in $T_c$ defined at the zero resistivity under these two applied fields is ~ $(11 \pm 1)$ (Fig.6c), indicating almost complete absence of field broadening. It is worth mentioning that such behavior is observed for conventional low-$T_c$ superconductors where the physics of flux pinning is dominated by quite different SC state parameters, e.g., $\xi$ is large and far less anisotropic, and $T_c$ itself is much lower. One possible reason for this small field broadening for the highly disordered sample could be due to the presence of unusually strong pinning in this compound. In fact, the ACS and the $\rho(T)$ measurements on S1 and S2 lend support for this assumption. The bulk of this 6%Zn sample contains significant parts which are poorly conducting and can act as a source of strong vortex pinning similar to the intrinsic pinning observed in high-$T_c$ cuprates when flux-flow in the poorly conducting c-direction is greatly blocked by the poorly conducting charge reservoir layers.

We summarise our findings regarding the PG in Fig.7. For comparison we have also shown $T^*(p)$ for a number of $Y_{1-x}Ca_xBa_2(Cu_{1-y}Zn_y)_3O_{7-\delta}$ compounds obtained from some of our previous studies[14,19,24]. The contrast in the evolution of $T^*(p)$ between $y > 0.05$ samples and all the other compounds is striking. For these very heavily disordered systems $T^*(p)$ is anomalous to say the least.

## DISCUSSION

The observed PG-like features for highly disordered $Y_{1-x}Ca_xBa_2(Cu_{1-y}Zn_y)_3O_{7-\delta}$ is surprising considering the widely held belief that the energy scale of the PG is determined by p, and $T^*(p)$ either falls below or merges with the $T_c(p)$ line in the OD side. In a previous magnetic susceptibility study, Cooper and Loram[34] observed a PG-like feature in fully oxygenated sintered 7%Zn-Y123, where superconductivity was almost completely suppressed with Zn. Vobornik *et al.* also observed a pseudogap in ARPES experiments for disordered optimally doped Bi-2212 single crystals[35]. For Bi-2212, in-plane disorder was introduced by electron irradiation and $T_c$ was reduced from 90K to 62K. $\rho_{ab}(T)$ was also measured but no high-T downturn in resistivity for this 62K optimally doped Bi-2212 was found[35]. In the light of present results, a possible reason might be that the sample was not "sufficiently" disordered, since we also did not see any anomalous effect in our 3%, 4%, and 5%Zn samples (where $T_c$ was suppressed by ~ 30K, ~ 40K, and ~ 46K respectively, near the optimum doping). Notice that all the previous observations of disorder induced PG were for samples that are slightly OD[34] and optimally doped respectively[35]. Here we have reported this effect over a fairly wide doping range from p = 0.205 to 0.11.

It is fair to say that there are two main effects of Zn substitution for Cu(2) in cuprates[4,19,28,36,37]: (i) Zn tends to localize carriers in the $CuO_2$ plane (depending on temperature and Zn content) and (ii) it gives rise to "*Curie-like magnetic moments*" possibly on the four neighboring Cu sites (in the presence of short-range antiferromagnetically correlated background). Recent $^{63}Cu$ NMR[38] and inelastic neutron scattering studies[39] suggest that Zn enhances the AF correlation in Y123. There is growing evidence that an AF background with short-range order does not exist in overdoped cuprates with p > 0.19 [4]. The situation in highly disordered samples could be different, especially if there is phase segregation, and only a very weak form of superconductivity is present. Under these circumstances Zn induced enhancement of short-range AF correlation might be a possibility for highly disordered $Y_{1-x}Ca_{1-x}Ba_2(Cu_{1-y}Zn_y)_3O_{7-\delta}$ irrespective of the hole content. It is very interesting to note that both the effects of Zn substitution (enhancement of short-range AF correlation or/and carrier localization in the $CuO_2$ plane) can be directly linked to the existence of the pseudogap. AF fluctuation has been held widely responsible for (or at least partly contributing to) the formation of the PG by various theoretical and experimental studies[4,5,40,41]. In another scenario, using the t-J model, strong Coulomb repulsion has been taken as responsible for both pseudogap and superconducting pairing[42]. In the VHS scenario[43], the magnitude of the pseudogap, similar to a "Coulomb-gap"[44] increases with increasing Coulombic repulsion. Disorder leads to localization of carriers and thus weakens the electrostatic screening, which, in turn effectively increases the Coulomb interaction. As Zn plays a similar role on carrier transport in the $CuO_2$ planes, it can give rise to a pseudogap in the VHS scenario.

It is hard to reconcile any precursor-pairing picture with this anomalous $T^*(p)$. On the other hand, scenarios in which the PG originates from correlations that are not related and compete with superconductivity, can possibly offer an explanation. In the competing correlations scenario $T^*(p)$ and $T_c(p)$ are detrimental to each other and the growth of one is accompanied by the decrease of the other (at least in the pure compounds). Kohno et al.[45] have in fact shown in their theoretical analysis that disorder can indeed strengthen one competing phase (AF) significantly over the other (e.g., superconductivity).

The abrupt appearance of this anomalous PG-like feature in resistivity as a function of disorder content is indicative of some sort of threshold mechanism in action, consistent with some recent theoretical studies[5,46]. For example, in the study by Monthoux[5] it was shown that as quasiparticle life-time becomes shorter (with increasing scattering by spin fluctuations), it starts to "feel" long-range magnetic order even when only short-range correlations are present. The quasiparticle mean-free path relative to the correlation length of the short-range AF order determines the size of the pseudogap. As the quasiparticle life-time decreases with Zn, it is possible that at a certain level of Zn in the $CuO_2$ plane the scattering rate becomes high enough to dominate over the more gradual p-dependent scattering. Using the Hubbard model, it was shown that a high scattering rate leads to the removal of low-energy spectral weight[46]. In this scenario the local Coulomb repulsion, U, is the important parameter. Heavy Zn substitution increases the scattering rate, localizes quasiparticles and effectively enhances U/t (where t is the nearest-neighbor hopping energy) globally. Once U > 8t, the pseudogap becomes insensitive to U and therefore, should not show significant p-dependence[46].

Roughly speaking, the mean distance between Zn impurities, $l_i$, in the $CuO_2$ plane is given by ~ $r/\sqrt{z_{pl}}$, where $z_{pl} = 3y/2$ and $r$ ~ 3.9Å is the distance between Cu atoms in the plane. For y = 0.06, $l_i$ ~ 13Å, close to the value of the in-plane superconducting coherence length, $\xi_{ab}$ ~ 15 Å. This should imply that once the distance between Zn atoms becomes comparable with the superconducting coherence length, abrupt changes in the various normal and superconducting state properties take place.

Another strong possibility is that this apparently anomalous PG is completely different and independent from its conventional counterpart. The systematic behavior of $T_c(p,y)$ (see Fig.1), irrespective of Zn content, supports this assumption. In this situation a conventional $T^*(p)$ should still exist for highly disordered samples. Here a much stronger resistive feature due to the anomalous PG could obscure the features associated with the conventional $T^*(p)$ which can be detected from the $\rho(T,p)$ for less disordered compounds.

## CONCLUSIONS

In summary, we have observed PG-like features in highly disordered $Y_{1-x}Ca_xBa_2(Cu_{1-x}Zn_x)_3O_{7-\delta}$ from resistivity measurements. Contrary to the situation for UD cuprates, the PG temperature showed almost no p-dependence over a large range of hole content. The results reported here lend support to the ARPES works on optimally doped Bi-2212, where spectroscopic features of the electron-irradiated compound looked similar to those of UD samples[35]. The ACS data of these heavily disordered compounds shows the superfluid density to be extremely low. Magneto-transport data provide a preliminary indication of strong pinning of vortices, possibly because significant part of the heavily disordered bulk remains poorly conducting at low temperature.

## ACKNOWLEDGEMENTS

We thank J.W. Loram and C. Panagopoulos for helpful comments and suggestions. SHN acknowledges the financial support from the Commonwealth Scholarship Commission (UK), Darwin College, Cambridge Philosophical Society, Lundgren Fund, and the Department of Physics Cambridge University. RSI acknowledges financial support from Trinity College, Cambridge and the Cambridge Commonwealth Trust (UK).

*Corresponding author. E-mail: shn21@cam.ac.uk

**Figure captions (S.H. Naqib *et al.*):**
**[Paper title: Anomalous pseudogap and …….]**

Fig.1. $T_c(p)$ of sintered $Y_{0.80}Ca_{0.20}Ba_2(Cu_{1-y}Zn_y)_3O_{7-\delta}$. y-values are shown. The straight line shows the shift in $p_{opt}$ with Zn.

Fig.2. Anomalous $T^*(p)$ of (a) sintered $Y_{0.80}Ca_{0.20}Ba_2(Cu_{0.94}Zn_{0.06})_3O_{7-\delta}$ (S1), (b) sintered $Y_{0.80}Ca_{0.20}Ba_2(Cu_{0.945}Zn_{0.055})_3O_{7-\delta}$ (S2), and (c) c-axis oriented thin film of $YBa_2(Cu_{0.93}Zn_{0.07})_3O_{7-\delta}$ [from ref.28]. Straight lines are drawn to locate $T^*$. The thick vertical lines show the p-independence of $T^*$.

Fig.3. Resistivity of (a) sintered $Y_{0.80}Ca_{0.20}Ba_2(Cu_{0.985}Zn_{0.015})_3O_{7-\delta}$, (b) sintered $Y_{0.80}Ca_{0.20}Ba_2(Cu_{0.96}Zn_{0.04})_3O_{7-\delta}$, and (c) c-axis oriented thin film of $Y_{0.95}Ca_{0.05}Ba_2Cu_3O_{7-\delta}$. Straight lines are drawn to locate the pseudogap temperature, $T^*$. Arrows mark the usual $T^*(p)$.

Fig.4. (a) Mass-normalized low-field ($H_{rms}$ = 0.1 Oe) ACS of sintered S1 as a function of p, (b) mass-normalized low-field ACS of sintered S2 as a function of p (c) field dependence of the ACS of sintered S1 (p = 0.205 ± 0.004), and (d) field dependence of the ACS of powdered S1 (p = 0.205 ± 0.004).

Fig.5. (a) Mass-normalized low-field ACS of sintered $Y_{0.80}Ca_{0.20}Ba_2(Cu_{0.96}Zn_{0.04})_3O_{7-\delta}$ as a function of p, (b) mass-normalized ACS of sintered $Y_{0.95}Ca_{0.05}Ba_2(Cu_{1-y}Zn_y)_3O_{7-\delta}$ with different amounts of Zn (p = 0.192 ± 0.005 for these samples), (c) field dependence of the ACS of sintered $Y_{0.90}Ca_{0.10}Ba_2(Cu_{0.97}Zn_{0.03})_3O_{7-\delta}$ (p = 0.196 ± 0.004), and (d) field dependence of the ACS of powdered $Y_{0.90}Ca_{0.10}Ba_2(Cu_{0.97}Zn_{0.03})_3O_{7-\delta}$ (p = 0.196 ± 0.004). The first inter-grain coupling temperature is shown in (c).

Fig.6. The magnetic field broadening of the resistive transition for (a) sintered $Y_{0.80}Ca_{0.30}Ba_2(Cu_{0.97}Zn_{0.03})_3O_{7-\delta}$ (p = 0.185 ± 0.004), straight lines are drawn to locate $T_{c-onset}$s, (b) c-axis oriented thin films of $Y_{0.95}Ca_{0.05}Ba_2(Cu_{0.98}Zn_{0.02})_3O_{7-\delta}$ (p = 0.168 ± 0.004), and (c) sintered S1 (p = 0.188 ± 0.004).

Fig.7. The T-p *phase diagram*. Un-filled symbols show the data from ref.24 (where using Zn and magnetic field to suppress $T_c$ we have been able to track $T^*(p)$ below $T_{c0}(p) = T_c(p, x = 0, y = 0)$). Filled symbols represent anomalous $T^*$. Straight lines are drawn as guide to the eye. The dashed line represents the $T_{c0}(p)$ of pure Y123 with $T_{cmax}$ = 93K. Sintered samples are denoted by "s" and tha c-axis thin films by "f".

Fig.1

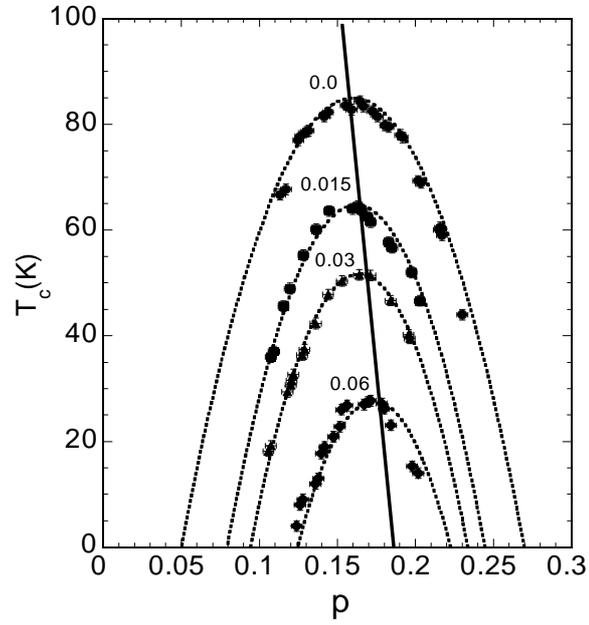

Figs.2

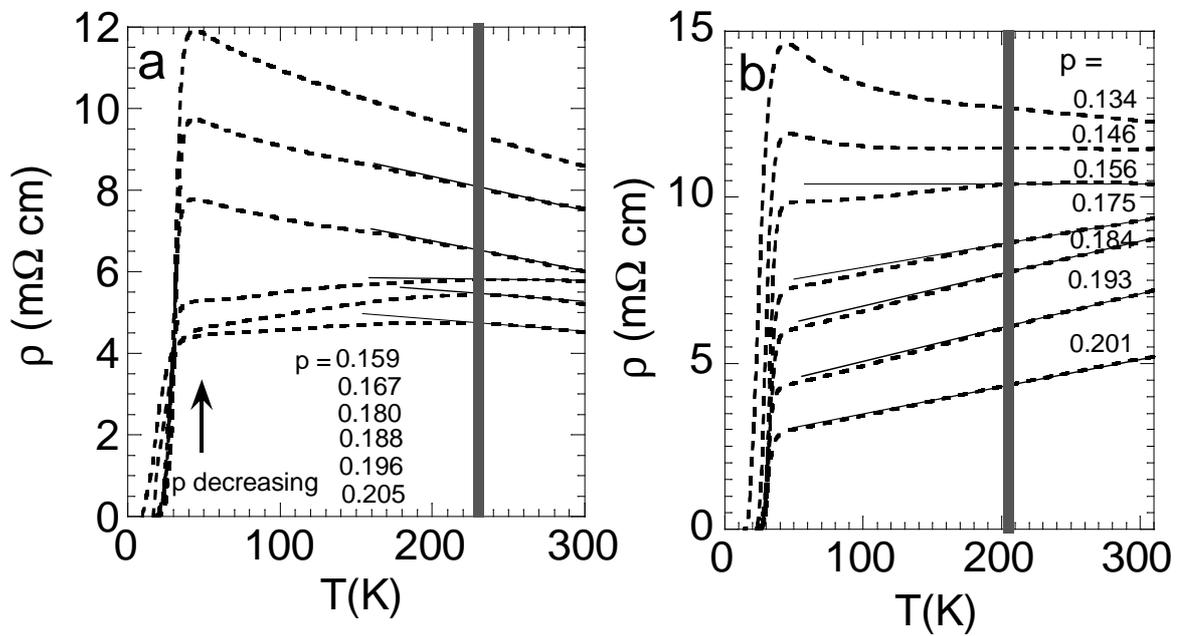

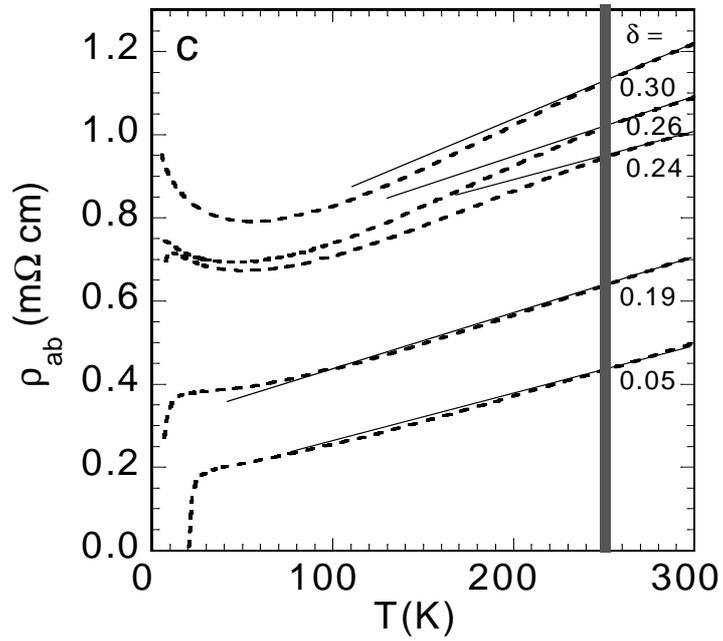

Figs.3

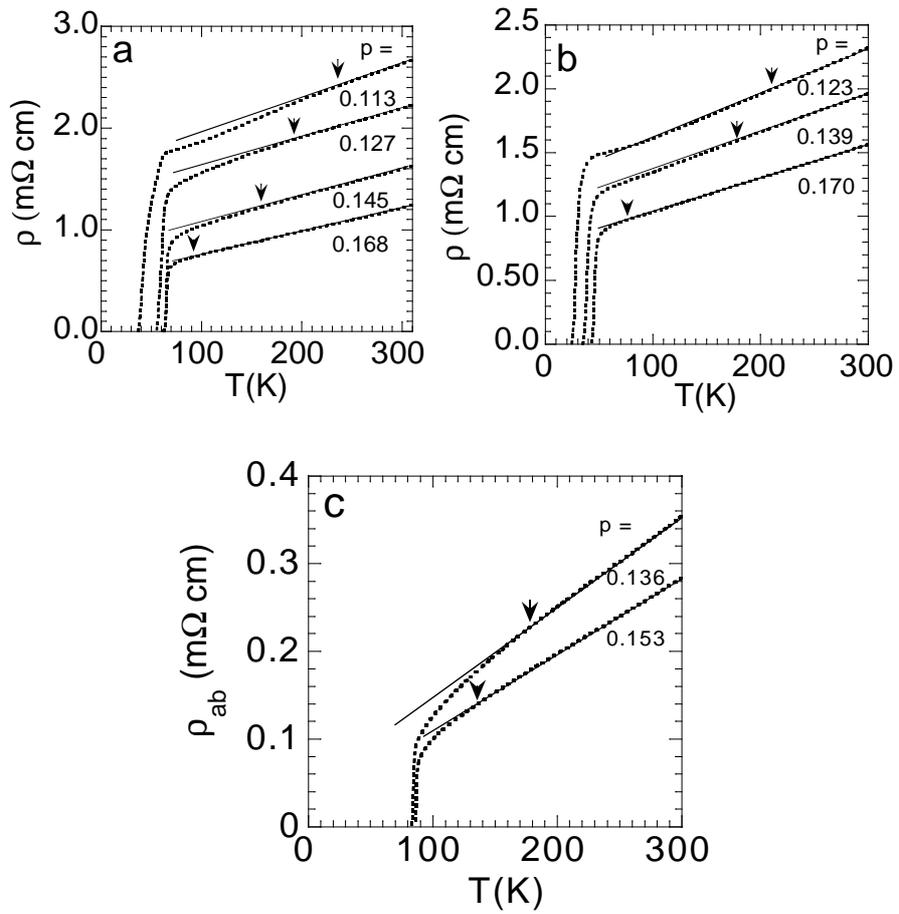

Figs.4

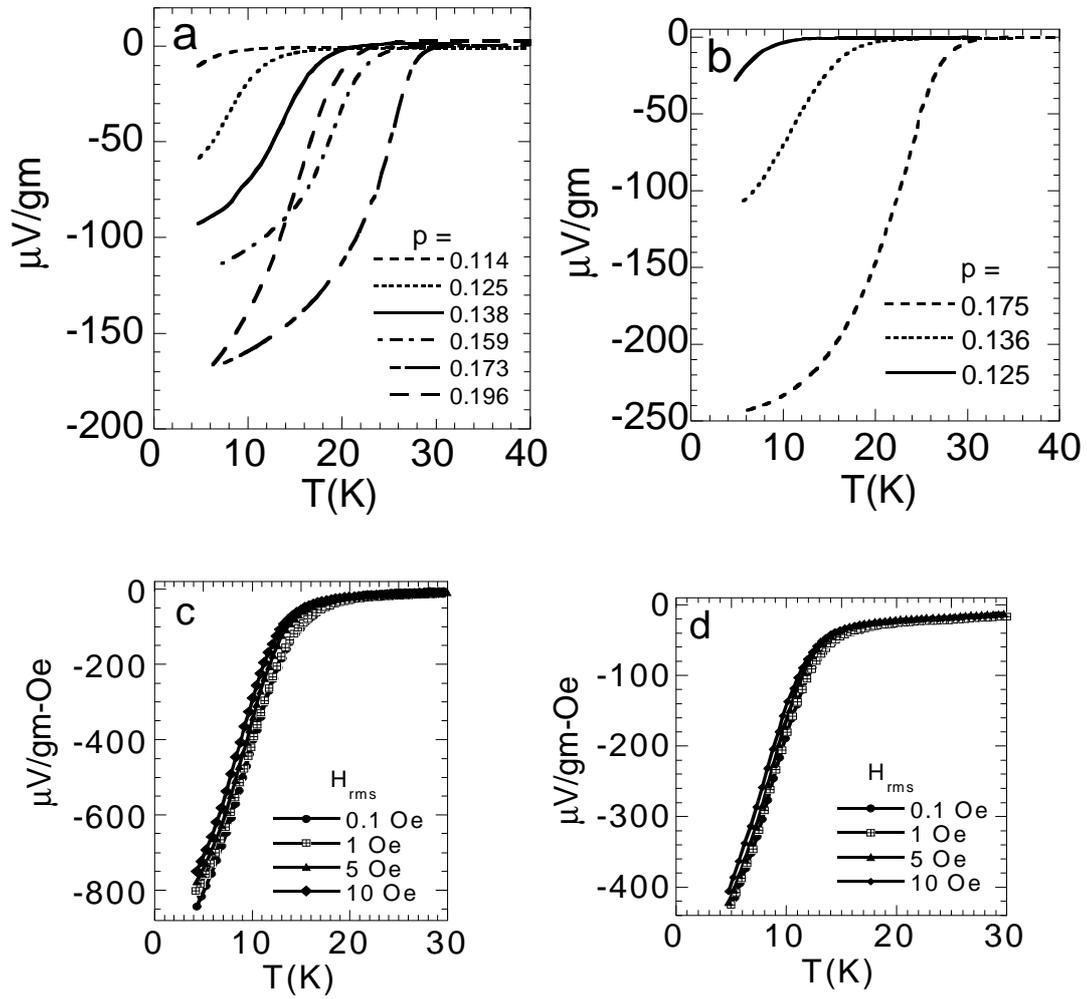

Figs.5

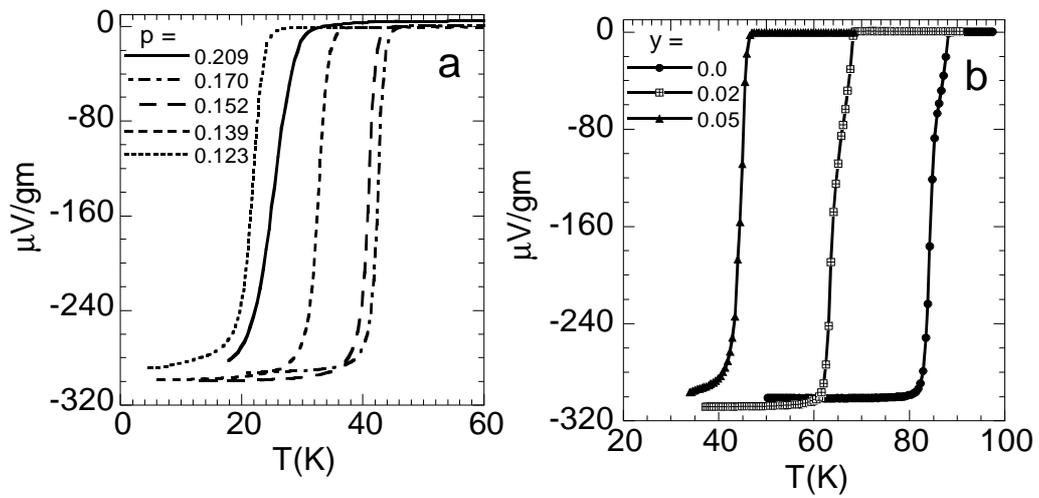

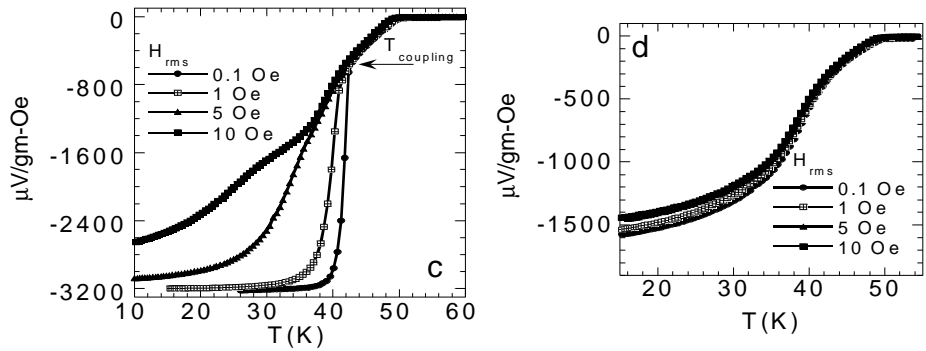

Figs.6

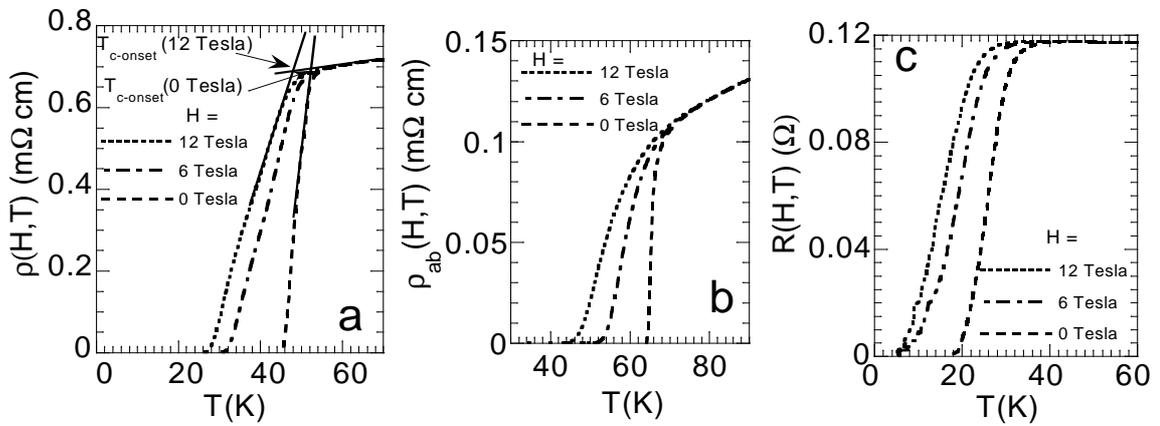

Fig.7

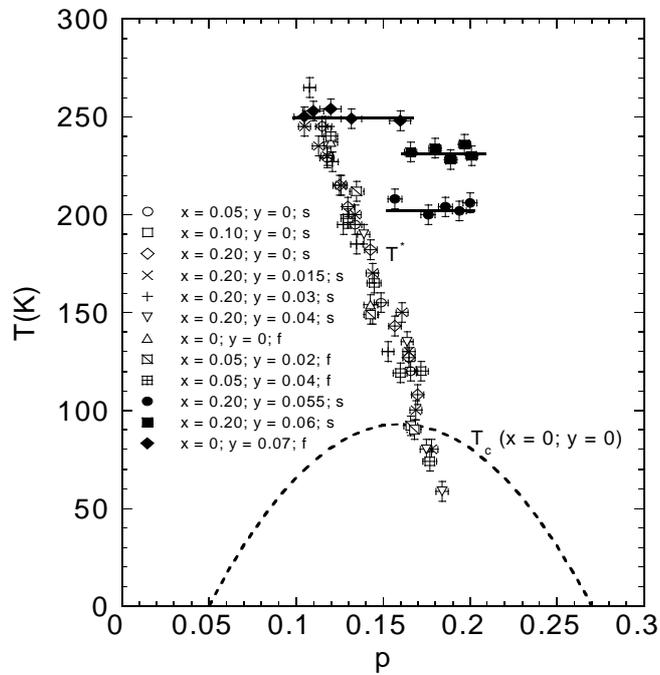